Email: sreetambhaduri@aol.com; ORCID: 0000-0002-5201-3976; Scopus ID: 57201682251

# The negative viscosity induces more disturbances in a flow

Sreetam Bhaduri

**Abstract:**

Negative viscosity seems to be an impossible parameter for any thermodynamic system. But for some special boundary conditions the viscosity of a fluid has apparently become negative, like for secondary flow of a fluid or in a plasma flow interacting with a dominant magnetic field. This work studied the effect of negative viscosity for a fluid flow over a cylinder. Four different viscosities are considered, in which the positive viscosities of Air and $CO_2$ has been considered at 300 K temperature and their negative pair of viscosities are considered in this work. The results show a vast difference in the vortex formation and pattern. General incompressible Navier Stokes equation has been employed for the analysis. The thermodynamic feasibility, vortex formation, variation of X direction velocity, variation of the VA factor and variation of drag coefficient has been studied subsequently in this work. SimFlow CFD software has been used in this work, which uses the OpenFOAM solver.

**Introduction:**

Viscosity is one of the important properties of a material in a fluid state. Viscosity provides an estimation of the resistance offered by the fluid layers to move relative to each other while in motion [1]. Therefore, viscosity provide information of frictional force in a fluid flow. As a result of the viscous force, the velocity of flow decreases. According to the second law of thermodynamics, all fluid requires to have positive viscosity [2, 3]. However, under some special boundary conditions an anisotropic fluid flow may show an overall negative viscosity [4, 5]. Sivashinsky et al. [4] mathematically demonstrated an isotropic external field when applied in a flow field in a certain direction, such that the effective viscosity became negative, a small perturbation in the magnitude of the applied field, result in a large-scale disturbance formation inside the flow. However, when the same field applied in all the directions equally, such that the effective viscosity remains positive, a small perturbation in the magnitude of the applied field dissipated quickly. The applied field may be a magnetic field or a secondary flow field. Gama et al. [6], calculated the wavenumber of these amplified resultant large-scale

Email: sreetambhaduri@aol.com; ORCID: 0000-0002-5201-3976; Scopus ID: 57201682251

disturbances are the square of that of the wave number of the small input perturbation. Therefore, the above literature confirms the effective viscosity can be positive or even negative also depending on the direction of the applied field. However, is it possible that the absolute viscosity of any certain fluid be negative according to the second law of thermodynamics? The next discussion will focus on this aspect and will find an expression of the rate of change of the entropy generation of the universe of an effective negative viscosity in a control volume system.

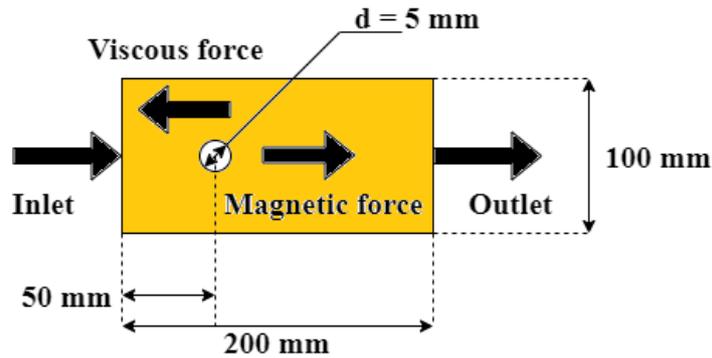

Fig.1. Problem description

Here, a benchmark problem of an external flow over an infinitely long cylinder has been considered for the analysis. This type of problems are better known as flow past cylinder. The flow over a cylinder shows a distinct feature of vortex shedding effect, which is widely known as Von-Karmann vortex street [7]. Therefore, in this work the effects on the vortex formation have been studied in detailed and the distinctive comparison has been done for the positively and negatively viscous flow. The Fig.1 shows schematic diagram of the problem. The generalized energy balance for any control volume system can be written as [8]

$$\dot{W}_{sh} + \dot{F}_l + \dot{m}\left[\Delta PE + \Delta KE + \frac{\Delta P}{\rho}\right] = 0 \qquad (1)$$

When the magnetic force taken into consideration, the equation (1) changes in to,

$$\dot{W}_{sh} + \dot{F}_l + \dot{m}\left[\Delta PE + \Delta KE + \frac{\Delta P}{\rho}\right] + \dot{F}_{em} = 0 \qquad (2)$$

According to this problem, $W_{sh} = 0$ and the $\Delta PE$ and $\Delta KE$ are neglected. Therefore, the equation (2) turns as the following,

$$\dot{F}_l + \frac{\dot{m}}{\rho}\Delta P + \dot{F}_{em} = 0 \qquad (3)$$

Email: sreetambhaduri@aol.com; ORCID: 0000-0002-5201-3976; Scopus ID: 57201682251

$$\therefore \dot{F}_l = -\frac{\dot{m}}{\rho}\Delta P - \dot{F}_{em} \tag{4}$$

The rate of dissipation of mechanical energy across the length of the flow domain,

$$\frac{\partial \dot{F}_l}{\partial x} = \frac{\dot{m}}{\rho}\left(-\frac{\partial P}{\partial x}\right) - \frac{\partial \dot{F}_{em}}{\partial x} \tag{5}$$

According to the second law of thermodynamics, it's evident [8],

$$\dot{S}_u = \dot{m}(\Delta s) > 0 \tag{6}$$

Also, for any control volume system [8],

$$Tds = dh - \frac{dP}{\rho} \tag{7}$$

Neglecting any heat transfer in this problem, we have $dh = 0$. Therefore, the equation (7) turned out to be,

$$\Delta s = -\frac{\Delta P}{\rho T} \tag{8}$$

Hence, the $\dot{S}_u = f(\Delta P)$. However, the mechanical energy dissipation can be directly related to the $\dot{S}_u$ where the effect of the magnetic force can also be taken in to consideration.

$$\dot{S}_u = \frac{\frac{\partial \dot{F}_l}{\partial x}}{TA_c} = \frac{\dot{m}}{\rho TA_c}\left(-\frac{\partial P}{\partial x}\right) - \frac{1}{TA_c}\frac{\partial \dot{F}_{em}}{\partial x} \tag{9}$$

When the magnetic force remains constant throughout the domain, the equation (9) turned out as follows,

$$\dot{S}_u = \frac{\dot{m}}{\rho TA_c}\left(-\frac{\partial P}{\partial x}\right) = \frac{\dot{m}}{\rho TA_c}\left(-\frac{\Delta P}{L}\right) \tag{10}$$

The equation (10) is the final form of the $\dot{S}_u$, which shows the criterion of thermodynamic feasibility of any fluid having negative viscosity. However, closely analyzing the equation (10), it's clear even with absence of the magnetic force if a fluid show $\frac{\Delta P}{L} < 0$, then the process will be thermodynamically feasible due to $\dot{S}_u > 0$. Hence, in this work the thermodynamic feasibility has to be satisfied in order to satisfy its practicality.

**Computational modeling:**

*Governing equation:*

Email: sreetambhaduri@aol.com; ORCID: 0000-0002-5201-3976; Scopus ID: 57201682251

The problem has been studied using the simplest incompressible Navier Stokes equation. The problem has been analyzed using three inlet Res i.e., Re = 100, 150 and 200. The negative viscosity concept has been derived as following way.

$$\frac{D\vec{u}}{Dt} = -\frac{1}{\rho}\nabla p + \frac{\mu}{\rho}\nabla^2\vec{u} \tag{11}$$

Equation (11) is the Navier Stokes equation for an incompressible flow with no influence on any external force [9]. But when the Lorentz appears, the equation (11) converts to MHD equation [10].

$$\frac{D\vec{u}}{Dt} = -\frac{1}{\rho}\nabla p + \frac{\mu}{\rho}\nabla^2\vec{u} + \frac{1}{\rho}(\vec{J} \times \vec{B}) \tag{12}$$

The equation (12) expresses the simplest form of MHD equation for an incompressible plasma flow with $\vec{J}$ as the current density and $\vec{B}$ as the magnetic field [11]. The equation (12) can be rewritten as,

$$\frac{D\vec{u}}{Dt} = -\frac{1}{\rho}\nabla p + \frac{\mu}{\rho}\nabla^2\vec{u} + \frac{1}{\rho}|\vec{J}||\vec{B}|sin\theta \tag{13}$$

When $180° \leq \theta \leq 270° \Rightarrow 0 \leq sin\theta \leq -1$. Therefore, the equation (13) will be transformed into the following form.

$$\frac{D\vec{u}}{Dt} = -\frac{1}{\rho}\nabla p + \frac{\mu}{\rho}\nabla^2\vec{u}, \text{ when } \theta = 180° \tag{14a}$$

$$\frac{D\vec{u}}{Dt} = -\frac{1}{\rho}\nabla p + \frac{\mu}{\rho}\nabla^2\vec{u} - \frac{\mu}{\rho}|\vec{J}||\vec{B}|, \text{ when } \theta = 270° \tag{14b}$$

Therefore, for $180° < \theta < 270°$, the equation (13) can be as follows,

$$\frac{D\vec{u}}{Dt} = -\frac{1}{\rho}\nabla p + \frac{\mu}{\rho}\nabla^2\vec{u} - \frac{\mu}{\rho}|\vec{J}||\vec{B}||sin\theta| \tag{15}$$

Now if, the magnitude of Lorentz force is greater than the viscous force, then mathematically the function $NG$ is,

$$NG = \frac{\mu}{\rho}\nabla^2\vec{u} - \frac{1}{\rho}|\vec{J}||\vec{B}||sin\theta| < 0 \Rightarrow NG = \zeta\nabla^2\vec{u} < 0$$

Therefore, the equation (15) can be rewritten as,

$$\frac{D\vec{u}}{Dt} = -\frac{1}{\rho}\nabla p + \zeta\nabla^2\vec{u} \tag{16}$$

Email: sreetambhaduri@aol.com; ORCID: 0000-0002-5201-3976; Scopus ID: 57201682251

The equation (16) shows an effective negative viscosity of the fluid and this equation has been used in this work.

*Mesh generation:*

The CFD domain has been discretized with hex dominant mesh, which is in OpenFOAM software well known as block mesh. According to the physics of flow past cylinder, it's highly important to have a dense all around, as well as behind the cylinder to achieve an accurate high-fidelity result of the vortex shedding. However, according the previous literature survey (Gama et al. [6]) clearly shows the small perturbation in the applied external field amplifies to square of the perturbation magnitude as a result of negative viscosity. Therefore, it is expected to have a high disturbance all across the flow domain. Due to this purpose, a refinement of level (n) 2 has been provided behind the cylinder and a circular region of radius 0.05 mm (Fig.2). The local grid size in the refined region is given by the following way [12],

$$Local\ grid\ size = \frac{Grid\ size}{2^n} \qquad (17)$$

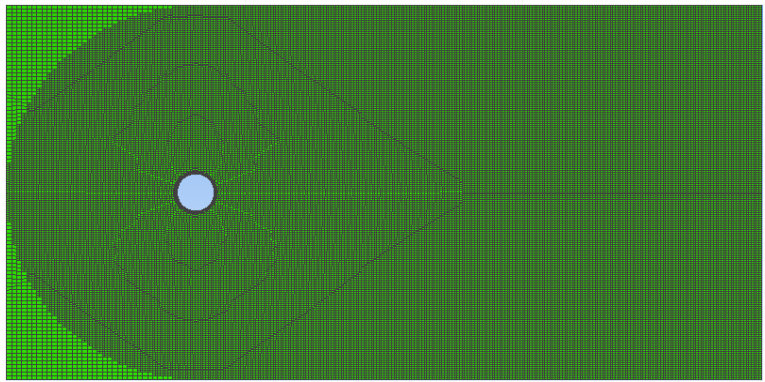

Fig.2. Meshed CFD domain

While grid generation, it is important to determine the proper grid size. This is performed by the Grid Independent Test. According to this test, if the results do not change with a small increase or decrease of the grid sizes w.r.t. an initial grid size, then that initial grid size is referred to as an optimal grid size. In this work the grid sizes varied as 0.9 mm, 1.2 mm, 1.5 mm, 1.8 mm and 2.1 mm. The Fig.3 shows the cross-sectional variation of X direction velocity at 0.9 mm distance from the center of cylinder along the direction of the flow. According to the Fig.3, it's clear that the variation of X direction velocity for different grid sizes are <1%, which is quite acceptable. Hence, the grid size of 0.9 mm has been adopted in this study. The local grid size in the refinement region is around 0.225 mm and the total number of nodes generated is around 60744 in the flow domain.

Email: sreetambhaduri@aol.com; ORCID: 0000-0002-5201-3976; Scopus ID: 57201682251

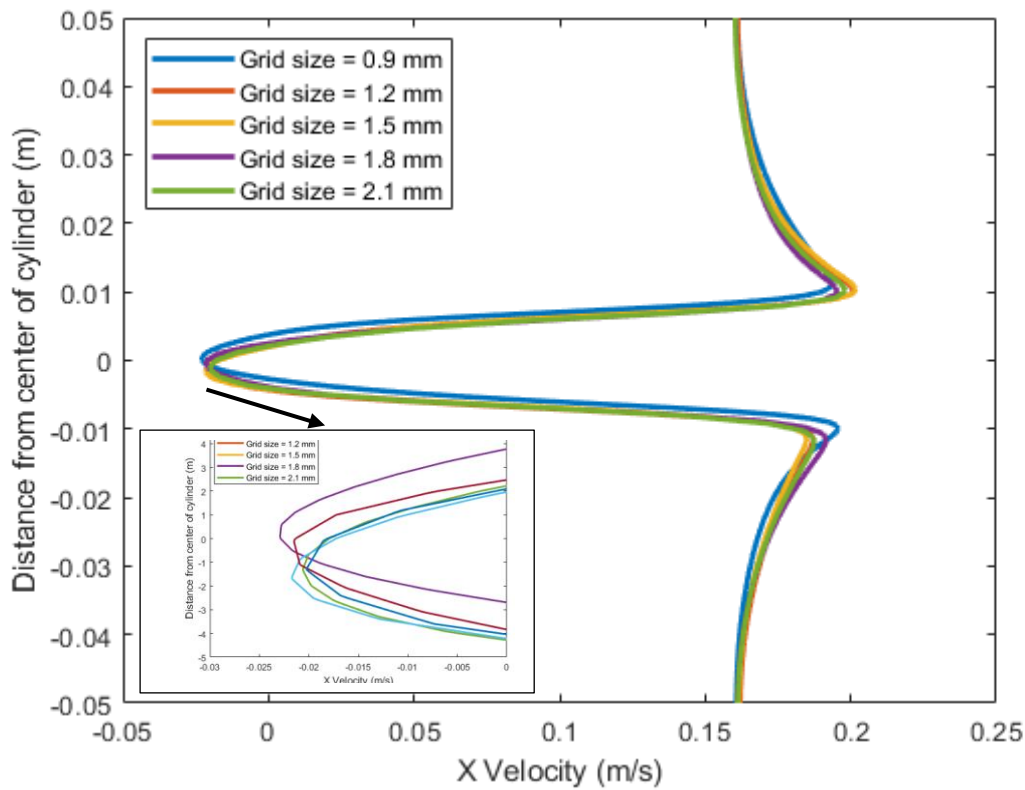

Fig.3. Variation of X direction velocity for different sizes of meshes

*Numerical solver:*

In this work, simple laminar flow has been adopted. Therefore, the icoFOAM solver has been used. The icoFOAM solver solves the incompressible laminar viscous flow using PISO algorithm [13]. The discretization of time scale and convection has been performed with Implicit Euler scheme and Linear-Upwind Stabilized Transport (LUST) scheme [14, 15]. The gradient is discretized using Central scheme to achieve a low truncation error. The pressure and velocity have been calculated with Geometric Algebraic Multi-Grid (GAMG) scheme and Preconditioned bi-conjugate (PBiCG) Stabilized scheme respectively to get a faster convergence [16, 17]. The residual of pressure and velocity has been taken as 1e-6.

*Boundary conditions:*

In this work, three different Reynolds numbers are considered at the inlet boundary conditions. They are Re = 100, Re = 150 and Re = 200. Two standard fluids Air and $CO_2$ are considered in this work along with isothermal condition. The kinematic viscosity of Air is $\zeta = 1.51E - 05 \, m^2/s^2$ and that of for the $CO_2$ is $\zeta = 7.61E - 06 \, m^2/s^2$. In order to study the effect of negative viscosity, the corresponding negative viscosities are considered as $\zeta = -1.51E -$

Email: sreetambhaduri@aol.com; ORCID: 0000-0002-5201-3976; Scopus ID: 57201682251

$05 \, m^2/s^2$ and $\zeta = -7.61E - 06 \, m^2/s^2$. Therefore, the four viscosities are $\zeta = 1.51E - 05 \, m^2/s^2$, $\zeta = 7.61E - 06 \, m^2/s^2$, $\zeta = -7.61E - 06 \, m^2/s^2$ and $\zeta = -1.51E - 05 \, m^2/s^2$. The total flow time has been taken as 5 s along with a timestep of 0.001 s. The OpenFOAM solver works with implicit solver, which eliminates the concept of CFL number [18].

**Results and discussions:**

Practically the viscosity of a fluid always positive. However, for some special cases the effective viscosity will become negative [4, 5]. On the other side, the second law of thermodynamics requires the viscosity to be positive [2, 3]. Therefore, as discussed earlier the thermodynamic feasibility of the problem must be studied initially. The equation (10) clearly shows the entropy generation for the universe is a strong function of pressure difference between the inlet and outlet of the domain. Therefore, to determine the entropy generation of the universe the pressure difference between the inlet and outlet has to be studied.

*Variation of the pressure difference and the rate entropy generation of the universe:*

The entropy generation of the universe is an important phenomenon while studying the thermodynamic feasibility of any process to occur in reality. The equation (10) determines the rate of entropy generation of the universe. Also, the equation (6) shows the criterion of feasibility of any process thermodynamically. Combining equation (6) and equation (10), it's clear that if the pressure difference is negative, means the process will be thermodynamically feasible. The Fig.4 shows the pressure difference between the inlet and outlet of the flow domain and the rate of entropy generation of the universe due to this pressure difference. It's clear that the pressure difference (Pressure at the outlet – Pressure at the inlet) is negative, therefore pressure loss happened in all the cases (Fig.4a, b and c). The pressure difference is negligible when the viscosity is positive due to low inlet Re. However, the pressure differences increased with increasing magnitude of negative viscosity except for Re = 150 due to increase in the vorticity magnitude (Fig.5, Fig.7 and Fig.9). According to equation (10) the pressure difference is directly proportional to the rate of entropy generation of the universe due to the flow. This is why, the rate of entropy generation of the universe due to the flow are positive for all the cases (Fig.4d, e and f) and follows a similar trend as of the pressure difference (Fig.4a, b and c). Hence, it is clear that in these cases are thermodynamically feasible according to the criterion of second law of thermodynamics (equation (6)).

| Pressure difference |
|---|

*Email: sreetambhaduri@aol.com; ORCID: 0000-0002-5201-3976; Scopus ID: 57201682251*

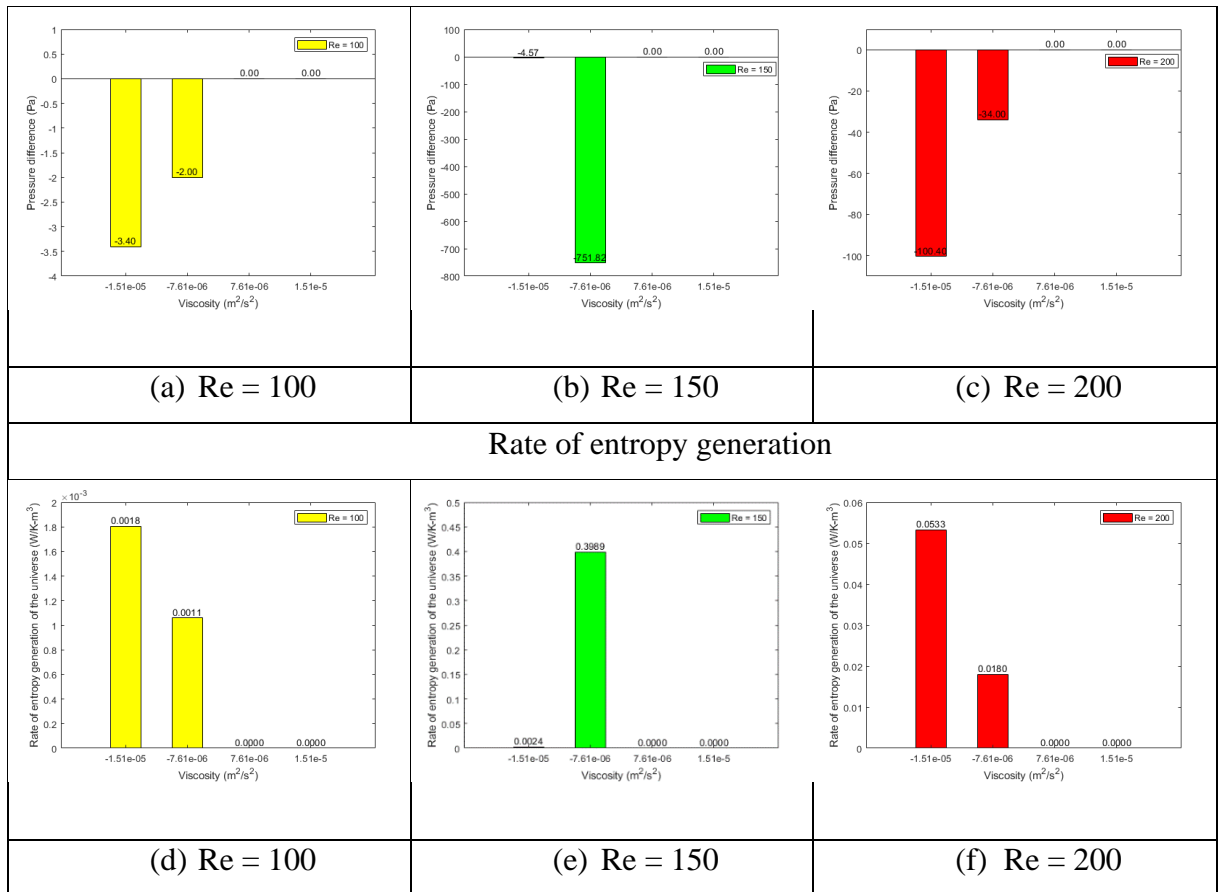

Fig.4. Comparison of pressure difference and rate of entropy generation for different viscosities

*Variation of vorticity magnitude with decreasing viscosity:*

The vorticity magnitude is an important flow parameter which quantifies the disturbances in a fluid flow. The positive viscosity has a tendency to dissipate any disturbances formed inside the flow by the frictional effect. However, negative magnitude of the fluid flow should have an opposite tendency, to increase the disturbances. Due to this the disturbances formed due to the flow obstruction by the cylinder, is expected to amplify inside the flow domain. The Fig.5 shows the spatial distribution of the vorticity magnitude in the flow with the inlet Re = 100 at 5s. In these figures it's clear that the vorticity magnitude is higher for the viscosity of $\zeta = 1.51E - 05 \, m^2/s^2$ than that of the $\zeta = 7.61E - 06 \, m^2/s^2$ (Fig.5a and b). This is because of decreasing magnitude of viscosity. However, the amplitude of vortex street decreased with decreasing magnitude of positive viscosity due to lesser flow separation for the $\zeta = 7.61E - 06 \, m^2/s^2$ than $\zeta = 1.51E - 05 \, m^2/s^2$ (Fig.5a and b). Analyzing the effect of negative magnitude of viscosity, it's closely following the expected outcome of amplification of disturbances formed. The vorticity magnitude increased with increase in the magnitude of negative viscosity from $\zeta = -7.61E - 06 \, m^2/s^2$ to $\zeta = -1.51E - 05 \, m^2/s^2$ (Fig.5c and d).

Email: sreetambhaduri@aol.com; ORCID: 0000-0002-5201-3976; Scopus ID: 57201682251

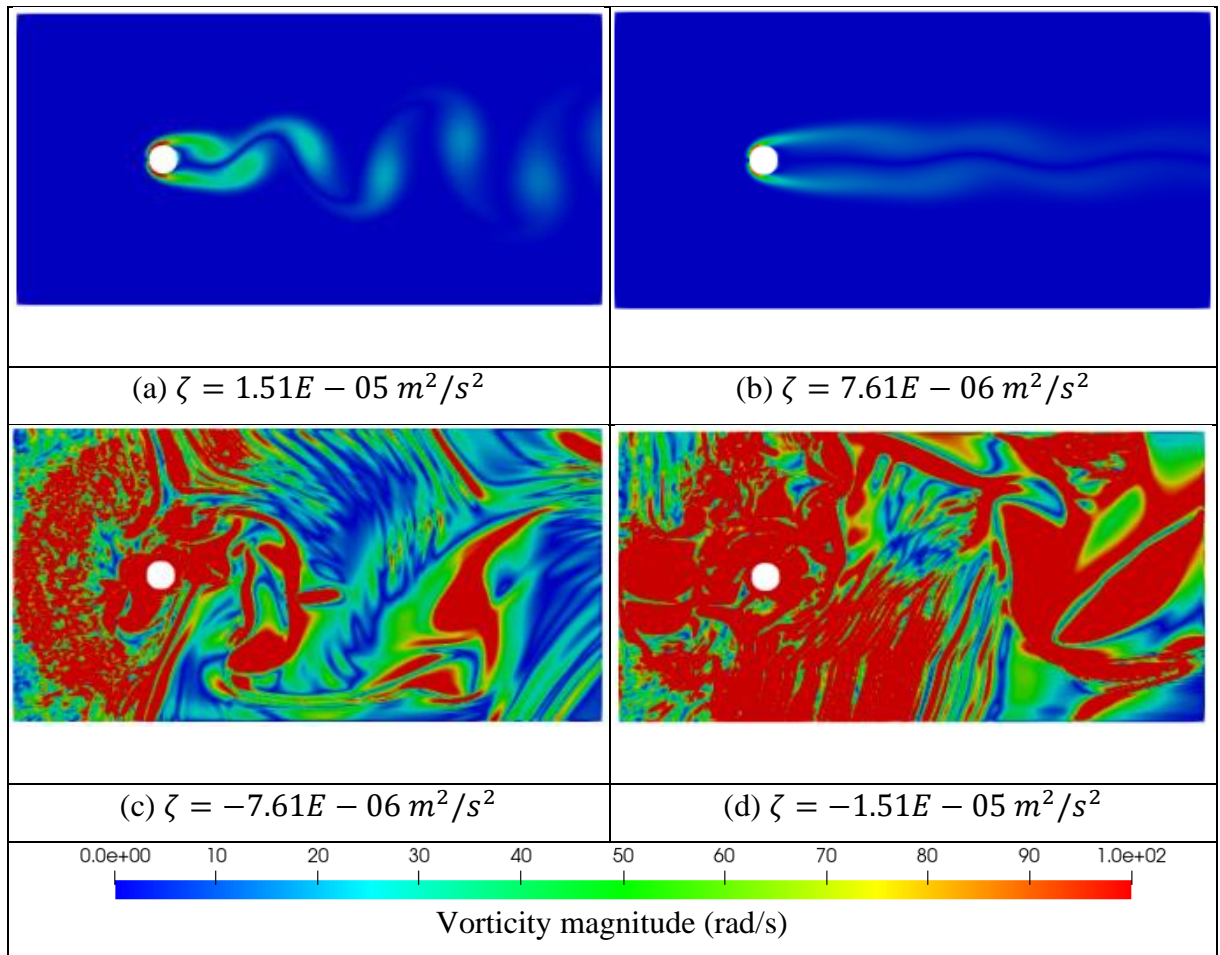

Fig.5. Spatial distribution of vorticity magnitude at 5s with Re = 100

The Fig.6 shows the comparison of the spatially averaged vorticity magnitude at 5s for the inlet Re = 100. Here the clear picture of decreasing vorticity magnitude can be observed with decreasing magnitude of positive viscosity, which is evident. However due to increase in the amplification of disturbances in the flow domain the vorticity magnitude increases with increasing magnitude of negative viscosity. The vorticity magnitude for the $\zeta = -1.51E - 05 \ m^2/s^2$ has been increased by 231 times w.r.t. $\zeta = 1.51E - 05 \ m^2/s^2$, and the vorticity magnitude for $\zeta = -7.61E - 06 \ m^2/s^2$ increased by 156 times w.r.t. $\zeta = 7.61E - 06 \ m^2/s^2$. In this case it's clear that the irrespective of the magnitude of viscosity, if the viscosity become negative in nature, the disturbances in the flow domain amplify increasingly.

*Email: sreetambhaduri@aol.com; ORCID: 0000-0002-5201-3976; Scopus ID: 57201682251*

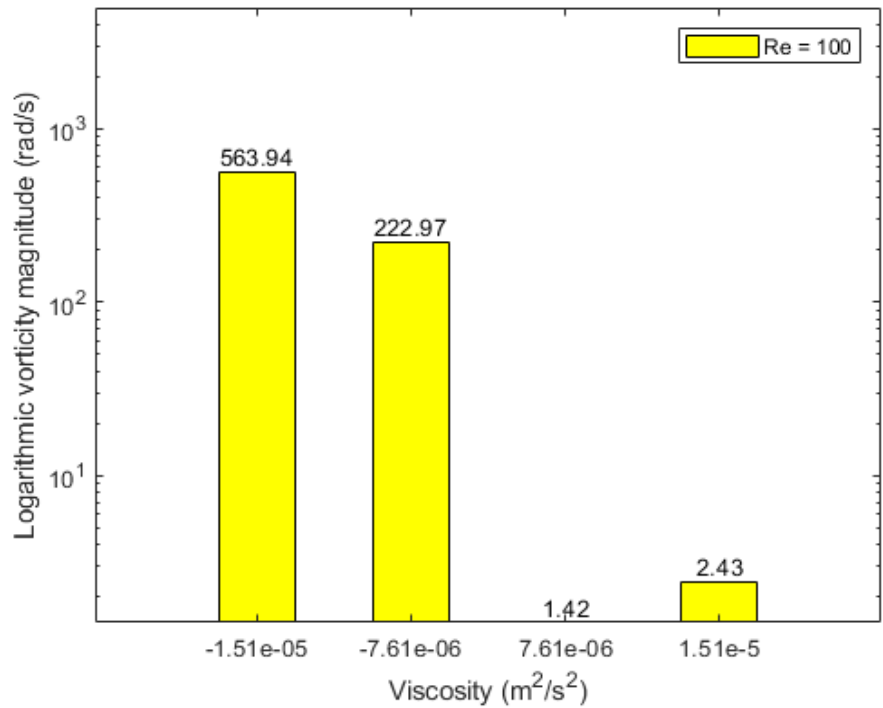

Fig.6. Spatially averaged vorticity magnitude at 5s with Re = 100

The Fig.7 shows the spatial distribution of vorticity magnitude at 5s for the inlet Re = 150. Due to increase of the magnitude of Re from 100 to 150, the vorticity magnitude increased for both positive and negative viscosities (Fig.5 and Fig.7). A similar trend of decreasing vorticity magnitude can be observed with decreasing magnitude of positive viscosity due to decreasing flow separation when the viscosity decreased from $\zeta = 1.51E - 05 \ m^2/s^2$ to $\zeta = 7.61E - 06 \ m^2/s^2$. However, in this case for Re = 150, the vorticity magnitude decreased with increasing magnitude of negative viscosity. This happened as a result of propagation of large-scale disturbances from the end of the flow domain towards the cylinder.

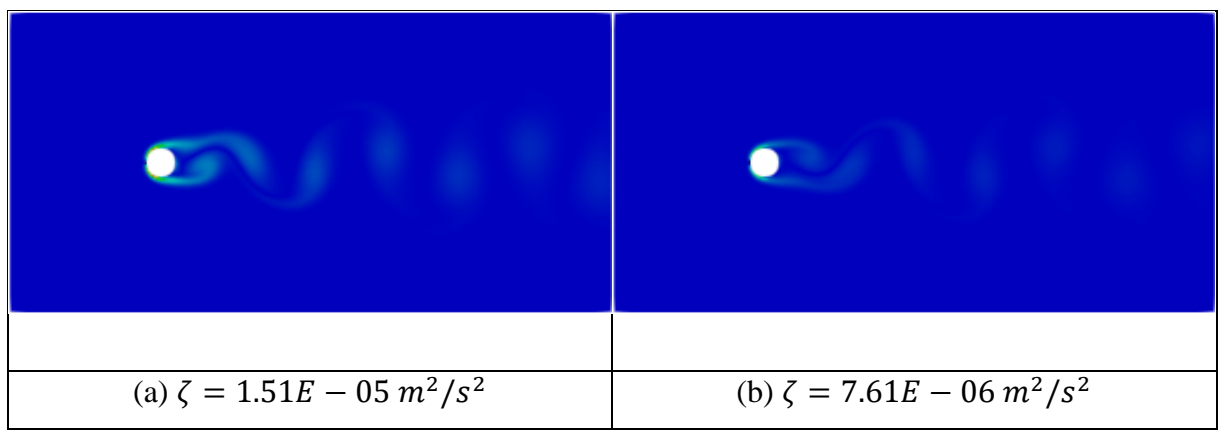

| (a) $\zeta = 1.51E - 05 \ m^2/s^2$ | (b) $\zeta = 7.61E - 06 \ m^2/s^2$ |

*Email: sreetambhaduri@aol.com; ORCID: 0000-0002-5201-3976; Scopus ID: 57201682251*

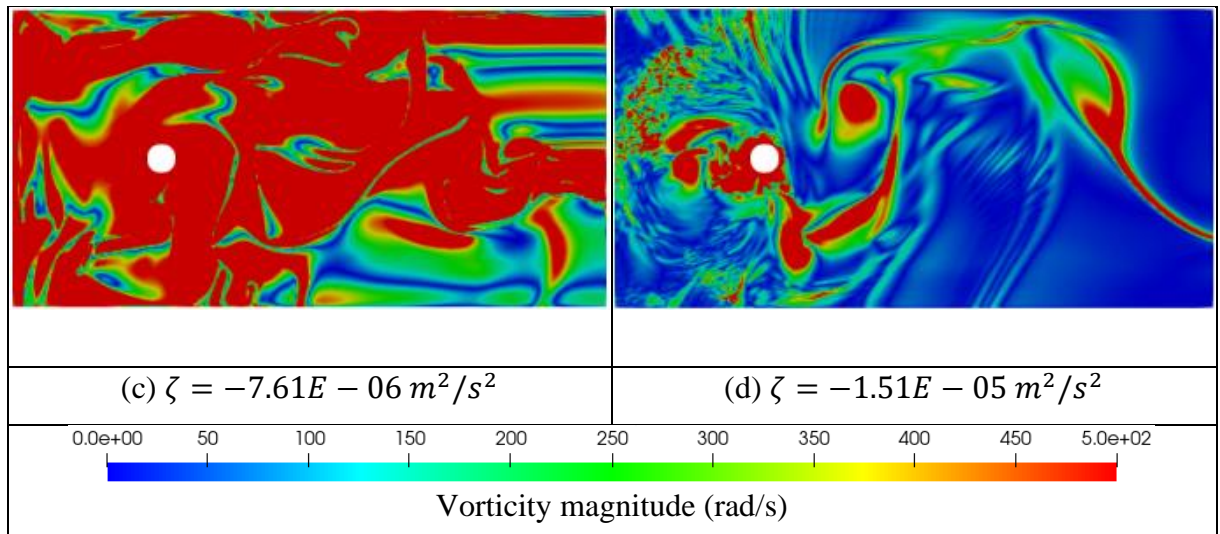

Fig.7. Spatial distribution of vorticity magnitude at 5s with Re = 150

The Fig.8 represents the comparison of spatially averaged vorticity magnitude at 5s for the inlet Re = 150. In this case also a similar trend of decreasing vorticity magnitude can be seen when the viscosity decreased from $\zeta = 1.51E - 05 \, m^2/s^2$ to $\zeta = 7.61E - 06 \, m^2/s^2$. The negative viscosity amplified the vorticity magnitude formed initially by the flow past cylinder. However, the vorticity magnitude for $\zeta = -1.51E - 05 \, m^2/s^2$ is 80.2% lower than that of $\zeta = -7.61E - 06 \, m^2/s^2$ as a result of forward propagation of large-scale disturbances from the end of flow domain. Even of this exception, the vorticity magnitude amplified by 130.8 times for $\zeta = -1.51E - 05 \, m^2/s^2$ w.r.t. that of $\zeta = 1.51E - 05 \, m^2/s^2$, and that of for the $\zeta = -7.61E - 06 \, m^2/s^2$ w.r.t. that of $\zeta = 7.61E - 06 \, m^2/s^2$ it is amplified by 1266.3 times respectively. In this case also, it is clear that the amplification of vorticity magnitude and the disturbance happens with the inclusion of negative viscosity than the positive viscosity.

Email: sreetambhaduri@aol.com; ORCID: 0000-0002-5201-3976; Scopus ID: 57201682251

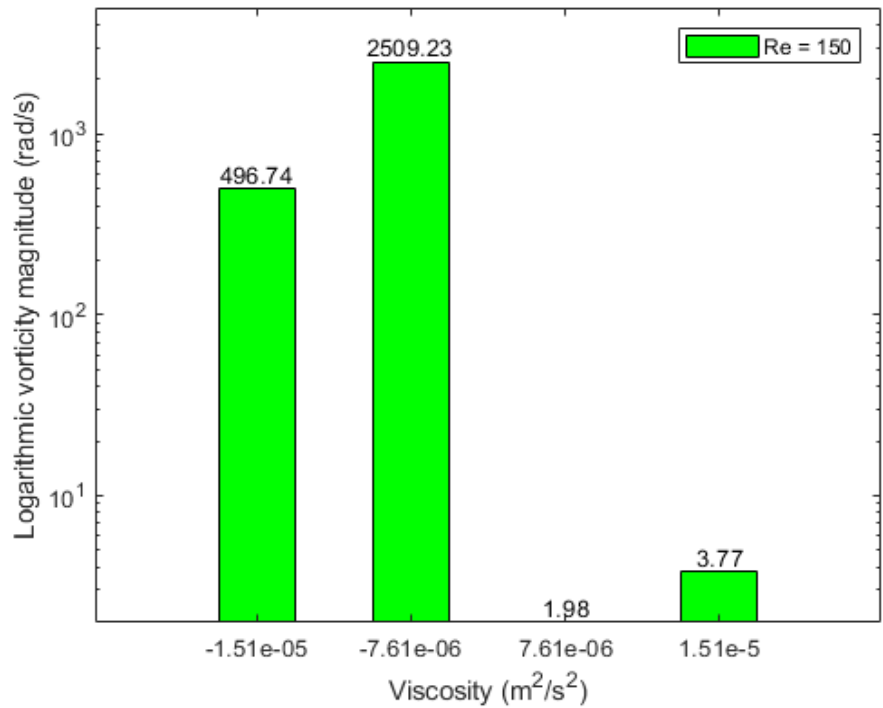

Fig.8. Spatially averaged vorticity magnitude at 5s with Re = 150

Similarly, the Fig.9 shows the spatial distribution of vorticity magnitude at 5s for the inlet Re = 200. Due to the decrease of viscosity from $\zeta = 1.51E - 05\ m^2/s^2$ to $\zeta = 7.61E - 06\ m^2/s^2$ the vorticity magnitude decreased as shown in Fig.9(a and b). The inclusion of negative viscosity again amplified the disturbances developed by the flow past cylinder in the entire domain (Fig.9c and d), which is absent for their positive counter parts (Fig.9a and b).

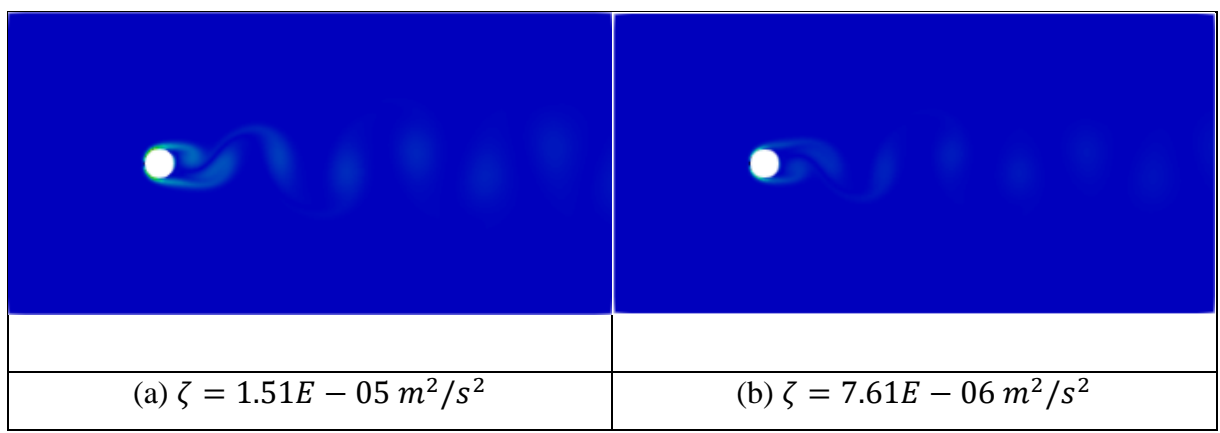

| (a) $\zeta = 1.51E - 05\ m^2/s^2$ | (b) $\zeta = 7.61E - 06\ m^2/s^2$ |

Email: sreetambhaduri@aol.com; ORCID: 0000-0002-5201-3976; Scopus ID: 57201682251

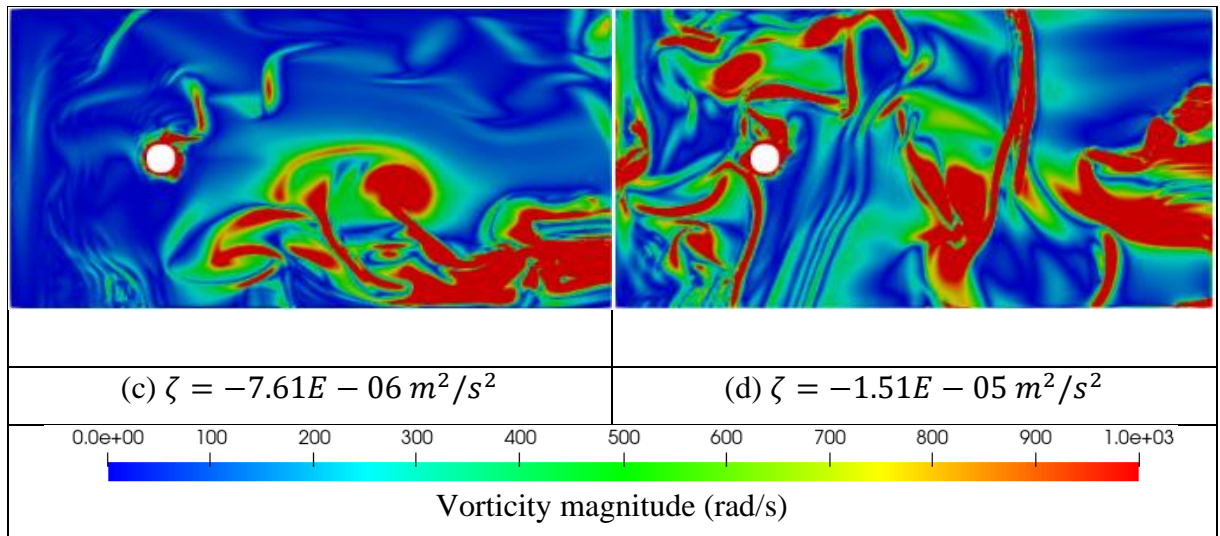

Fig.9. Spatial distribution of vorticity magnitude at 5s with Re = 200

The above spatial distribution for Re = 200 cannot confirm whether the increase of the magnitude of the negative viscosity increases the vorticity magnitude i.e., induces more disturbances in the domain. The Fig.10 clearly shows the comparison of spatially averaged vorticity magnitude at 5s for the inlet Re = 200. Following the Fig.10, it's clear that the vorticity magnitude decreased by 47.5% with decreasing positive viscosity and the vorticity magnitude increased by 68.33% with increasing magnitude of negative viscosity due to amplification of vorticity formed by the flow past cylinder. This trend is highly similar with the Re = 100 (Fig.6). The vorticity magnitude for $\zeta = -1.51E - 05\ m^2/s^2$ increased around 204.2 times w.r.t. $\zeta = 1.51E - 05\ m^2/s^2$. The vorticity magnitude for $\zeta = -7.61E - 06\ m^2/s^2$ increased around 231.5 times w.r.t. $\zeta = 7.61E - 06\ m^2/s^2$. So, in this case also, it's inevitable that the inclusion of negative viscosity develops more disturbances by amplifying the initial disturbances formed by the cylinder.

*Email:* <u>sreetambhaduri@aol.com</u>; *ORCID:* <u>0000-0002-5201-3976</u>; *Scopus ID:* <u>57201682251</u>

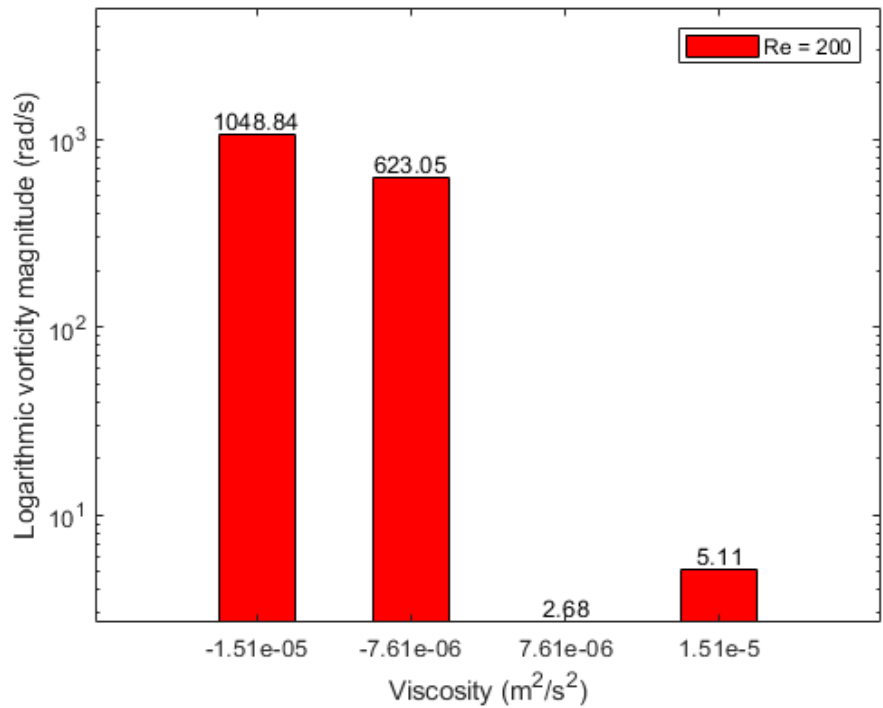

Fig.10. Spatially averaged vorticity magnitude at 5s with Re = 200

Therefore, the above discussion concludes that the negative viscosity amplifies the disturbances initially formed by the cylinder and the magnitude of vorticity generally keep increasing with increasing magnitude of the negative viscosity. Hence, the negative viscosity shows a viscous stimulating effect instead of the viscous dissipation effect which is predominantly existing in the positively viscous flow. The vorticity magnitude increased with increasing inlet Re from 100 to 200 for the positive viscosities. This is highly predictable due to increasing flow velocity. However, for the $\zeta = -7.61E - 06 \; m^2/s^2$ the vorticity magnitude is highest when the Re = 150 followed by Re = 200 and then Re = 100 due to the existence of forward propagating large-scale propagation for Re = 150. Also, for the $\zeta = -1.51E - 05 \; m^2/s^2$ the vorticity magnitude is the highest when the Re = 200 followed by Re = 100 and then Re = 150 as a result of varying of inlet Re.

*Variation of flow velocity across the cross section at 9 mm from the center of the cylinder along the flow direction:*

The previous discussion provides an insight of variation of the flow pattern with varying viscosities. However, no information of the magnitude of the velocity of the flow can be found. The following discussion focuses on the magnitude of the X direction velocity inside the flow at different time with varying viscosity. In the following discussion the point of inflection will

Email: sreetambhaduri@aol.com; ORCID: 0000-0002-5201-3976; Scopus ID: 57201682251

be an important parameter which determines the point where the X direction velocity changes the direction from positive to negative and vice versa. Also, to parametrize the number of sharp changes in the slope of the velocity variation across the cross section, the number of critical points is analyzed. They are mathematically defined as follows,

$$CP = (V, y) \; where, \left(\frac{dV}{dy}\right)_i = 0; \left(\frac{dV}{dy}\right)_{i+1} \neq 0 \; and \; \left(\frac{dV}{dy}\right)_{i-1} \neq 0 \quad (18)$$

Here $i$ = location of a point on the velocity curve.

Another important parameter to study is the ratio of terminal (maximum or minimum) X direction velocity across 9 mm distance from the center of cylinder to the inlet X direction velocity. The parameter has been named as the Velocity Amplification (VA) factor. The positive viscosity reduces the X direction velocity due to viscous dissipation effect and slightly increases the X direction velocity due to the curvature of the cylinder. However, the negative viscosity should increase the X direction velocity by the action of viscous stimulation effect. These are the requirements to define the VA factor. Mathematically the VA factor can be defined as follows,

$$VA \; factor = \frac{(V_x)_{max}}{V_{inlet}} \; or \; \frac{(V_x)_{min}}{V_{inlet}} \quad (19)$$

The further discussion will represent the inflection point as IP and the critical point as CP. The Fig.11 shows the comparison of the variation of X direction velocity across the cross-section at 9 mm distance from the center of the cylinder along the direction of flow for Re = 100. Positively viscous flow past cylinder generally obstruct flow just behind the cylinder surface, and develop vortices due to flow separation. Moreover, the obstruction by the cylinder, a sudden decrease of the velocity can be observed just behind the cylinder (diametrically). Due to these reasons, the X direction velocity corresponds to 3 CPs and 2 IPs (Fig.11a and b). However, due to viscous stimulation effect of the negative viscosity, the vorticity magnitude increased as seen from Fig.5(c and d). As a result, it's expected to have a highly fluctuating X direction velocity in these cases. A similar result can be observed for the negatively viscous flow in Fig.11(c and d). The number of CPs are 11 and the number of IPs are 6 for the $\zeta = -7.61E - 06 \; m^2/s^2$. The number of CPs are 10 and the number of IPs are 2 for the $\zeta = -1.51E - 05 \; m^2/s^2$. The number CPs and IPs reduced with increasing magnitude of negative viscosities because of increased inlet velocity for maintaining a constant inlet Re = 100.

Email: sreetambhaduri@aol.com; ORCID: 0000-0002-5201-3976; Scopus ID: 57201682251

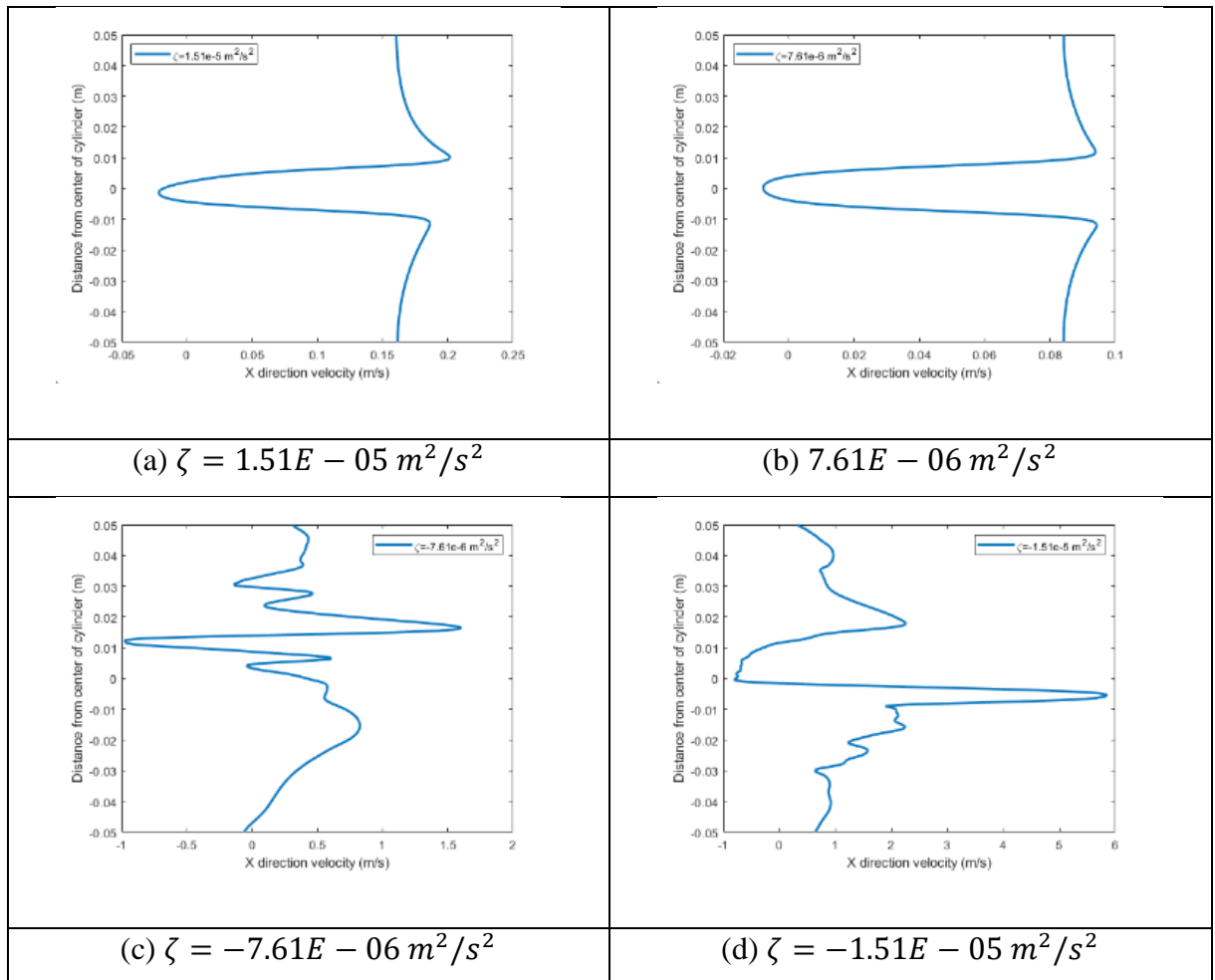

Fig.11. Cross-sectional variation of X direction velocity at 9 mm distance from the center of cylinder in the direction along the flow at 5s with Re = 100

The Fig.12 shows the comparison of the VA factor for different viscosities for the inlet Re = 100. The positive VA factor and negative VA factor decreased by 7.46% and 28.57% with decreasing viscosity from $\zeta = 1.51E - 05 \; m^2/s^2$ to $\zeta = 7.61E - 06 \; m^2/s^2$ due to decreasing viscous resistance. The positive VA factor increased by 83.78% with increasing magnitude of negative viscosity from $\zeta = -7.61E - 06 \; m^2/s^2$ to $\zeta = -1.51E - 05 \; m^2/s^2$ as a result of increasing viscous stimulation effect. However, the negative VA factor decreased by 58.3% with increasing magnitude of negative viscosity from $\zeta = -7.61E - 06 \; m^2/s^2$ to $\zeta = -1.51E - 05 \; m^2/s^2$ for similar reason of increasing viscous stimulation effect.

*Email: sreetambhaduri@aol.com; ORCID: 0000-0002-5201-3976; Scopus ID: 57201682251*

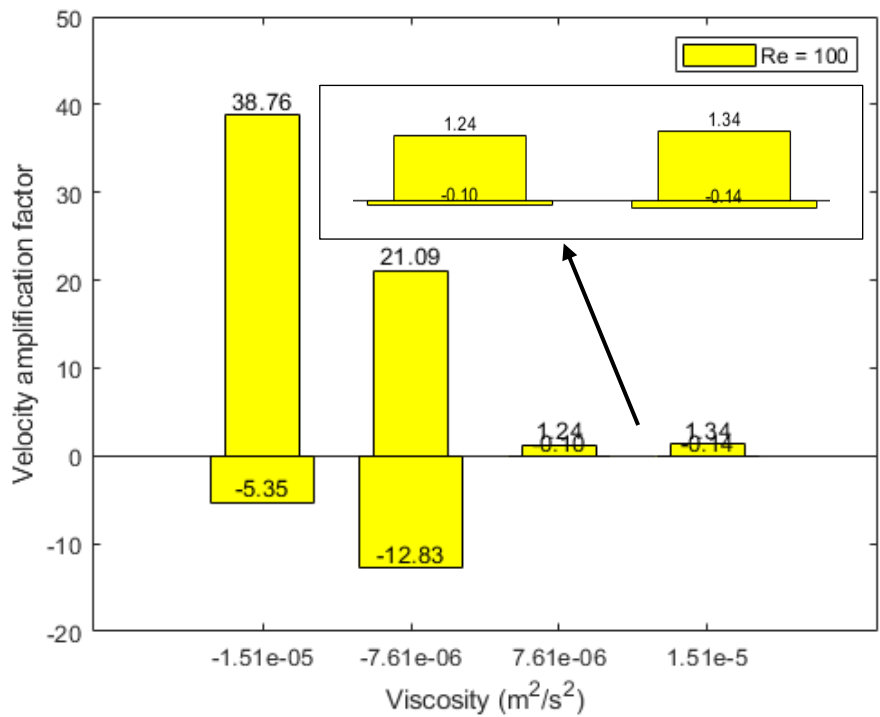

Fig.12. Comparison of VA factor across cross section at 9 mm distance from center of cylinder w.r.t. the inlet velocity for different viscosities at Re = 100

The Fig.13 shows the comparison of variation of X direction velocity at 9 mm distance from the center of cylinder at 5s for the inlet Re = 150. The number of CPs and IPs for $\zeta = 1.51E - 05 \, m^2/s^2$ and $\zeta = 7.61E - 06 \, m^2/s^2$ are 3 and 2 respectively. However, due to viscous stimulation effect the number of CPs and IPs for $\zeta = -7.61E - 06 \, m^2/s^2$ are 6 and 2, and that of for $\zeta = -1.51E - 05 \, m^2/s^2$ are 8 and 2 respectively. The number of IPs remain same with increasing magnitude of negative viscosity due to increased inlet to Re = 150 than that of Re = 100 (Fig.11). The number of CPs increased with increasing magnitude of negative viscosity due to the cumulative effect of increased inlet velocity to maintain constant inlet Re = 150 and increased viscous stimulation effect.

Email: sreetambhaduri@aol.com; ORCID: 0000-0002-5201-3976; Scopus ID: 57201682251

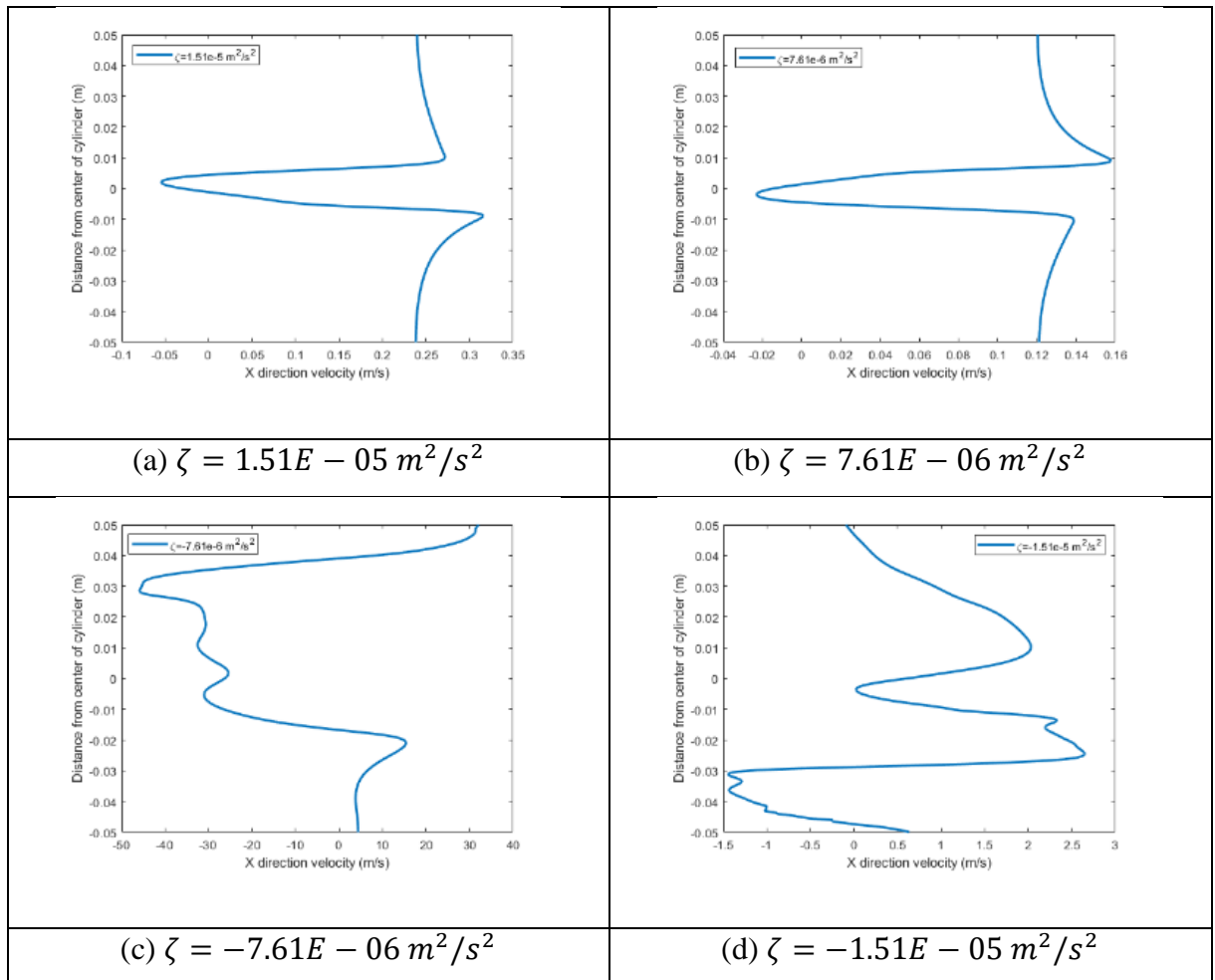

(a) $\zeta = 1.51E - 05 \; m^2/s^2$

(b) $\zeta = 7.61E - 06 \; m^2/s^2$

(c) $\zeta = -7.61E - 06 \; m^2/s^2$

(d) $\zeta = -1.51E - 05 \; m^2/s^2$

Fig.13. Cross-sectional variation of X direction velocity at 9 mm distance from the center of cylinder in the direction along the flow at 5s with Re = 150

The Fig.14 shows the comparison of the VA factor for different viscosities for the inlet Re = 150. Similarly, the positive VA factor and negative VA factor decreased by 0.72% and 16.67% with decreasing viscosity from $\zeta = 1.51E - 05 \; m^2/s^2$ to $\zeta = 7.61E - 06 \; m^2/s^2$ due to decreasing viscous resistance. The positive VA factor and negative VA factor decreased by 95.83% and 99.9% with increasing magnitude of negative viscosity from $\zeta = -7.61E - 06 \; m^2/s^2$ to $\zeta = -1.51E - 05 \; m^2/s^2$ due to the forward propagation of large-scale disturbances from the end of the flow domain.

Email: sreetambhaduri@aol.com; ORCID: 0000-0002-5201-3976; Scopus ID: 57201682251

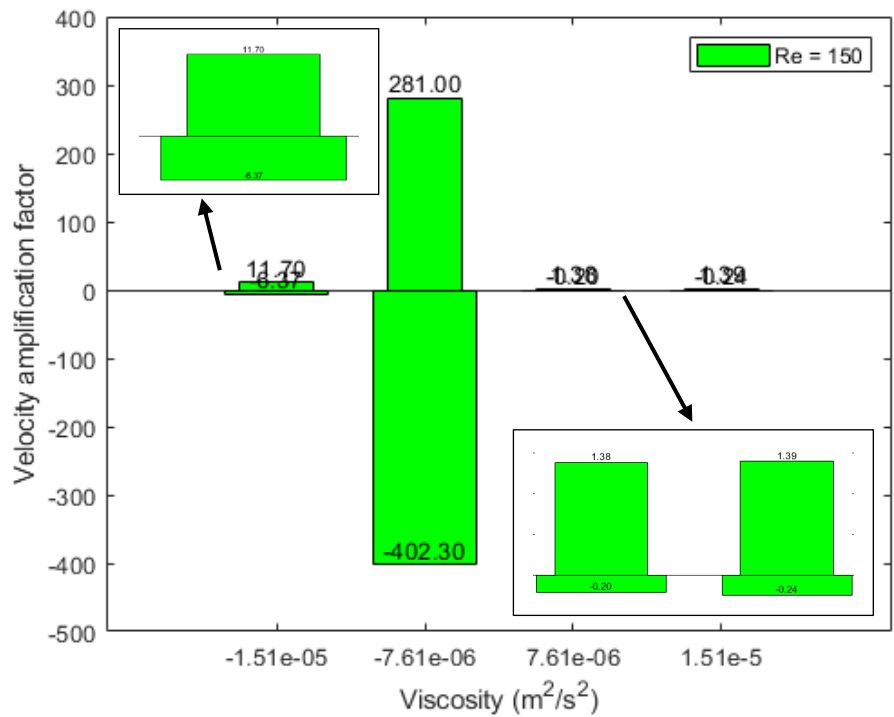

Fig.14. Comparison of VA factor across cross section at 9 mm distance from center of cylinder w.r.t. the inlet velocity for different viscosities at Re = 150

The Fig.15 shows the comparison of variation of X direction velocity at 9 mm distance from the center of cylinder at 5s for the inlet Re = 200. There 3 CPs and 2 IPs are present for positive viscosities as shown in Fig.15(a and b). As a result of viscous stimulation effect for the negative viscosities, a high fluctuation is noticeable. The number of CPs and IPs for $\zeta = -7.61E - 06 \ m^2/s^2$ and $\zeta = -1.51E - 05 \ m^2/s^2$ are 6, 3 (Fig.15c) and 4, 3 (Fig.15d) respectively as a result of increased inlet X direction velocity in order to maintain a constant inlet Re = 200.

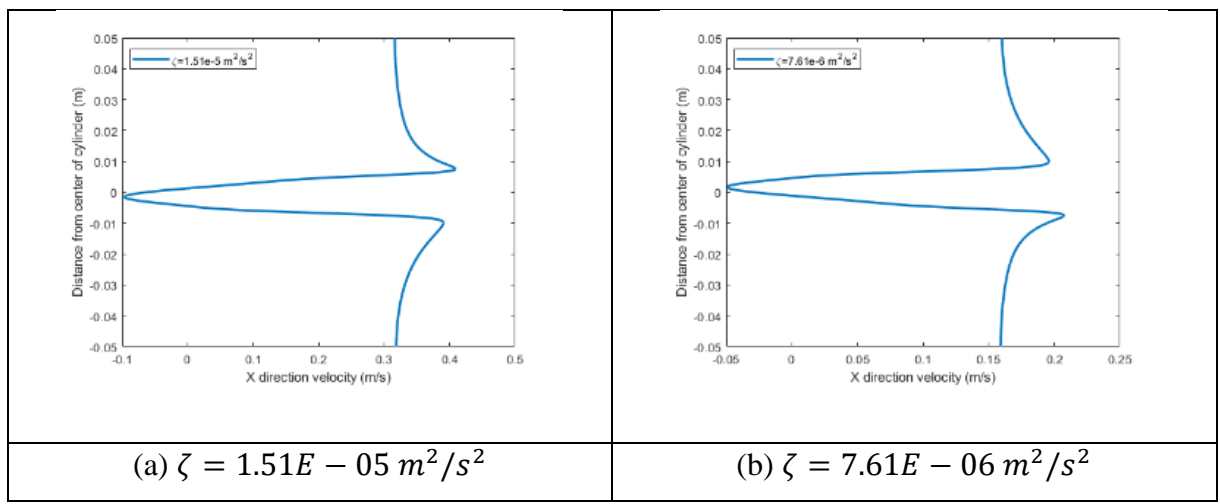

| (a) $\zeta = 1.51E - 05 \ m^2/s^2$ | (b) $\zeta = 7.61E - 06 \ m^2/s^2$ |

Email: sreetambhaduri@aol.com; ORCID: 0000-0002-5201-3976; Scopus ID: 57201682251

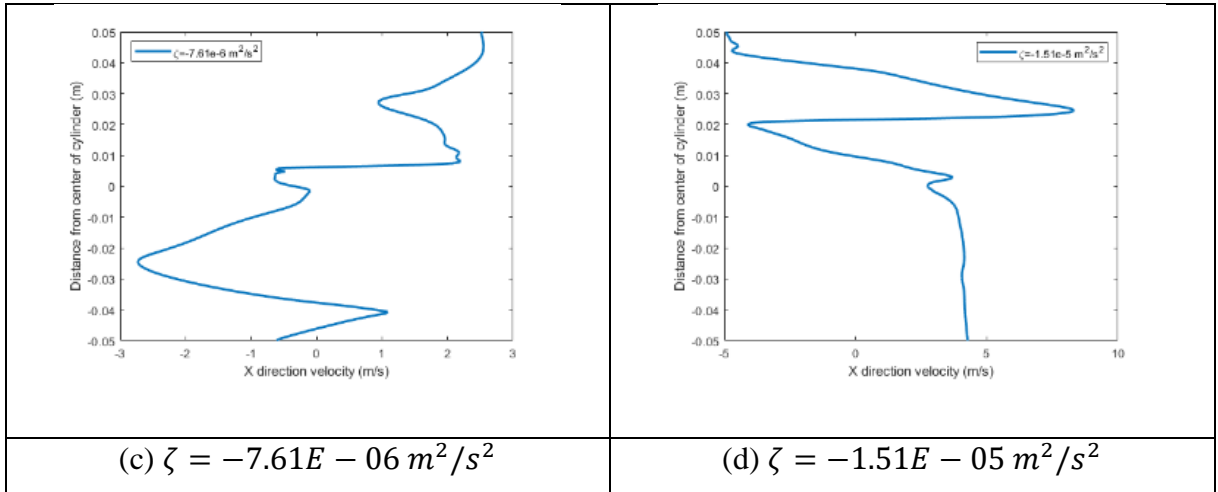

| (c) $\zeta = -7.61E - 06\ m^2/s^2$ | (d) $\zeta = -1.51E - 05\ m^2/s^2$ |

Fig.15. Cross-sectional variation of X direction velocity at 9 mm distance from the center of cylinder in the direction along the flow at 5s with Re = 200

The Fig.16 shows the comparison of the VA factor for different viscosities for the inlet Re = 200. The positive VA factor and negative VA factor are almost same with decreasing viscosity from $\zeta = 1.51E - 05\ m^2/s^2$ to $\zeta = 7.61E - 06\ m^2/s^2$ due to increased inlet Re than the previous cases (Fig.12 and Fig.14). The positive VA factor increased by 64.22% with increasing magnitude of negative viscosity from $\zeta = -7.61E - 06\ m^2/s^2$ to $\zeta = -1.51E - 05\ m^2/s^2$ due to increasing viscous stimulation effect. However, the negative VA factor decreased by 9.42% with increasing magnitude of negative viscosity from $\zeta = -7.61E - 06\ m^2/s^2$ to $\zeta = -1.51E - 05\ m^2/s^2$ due to increasing viscous stimulation effect.

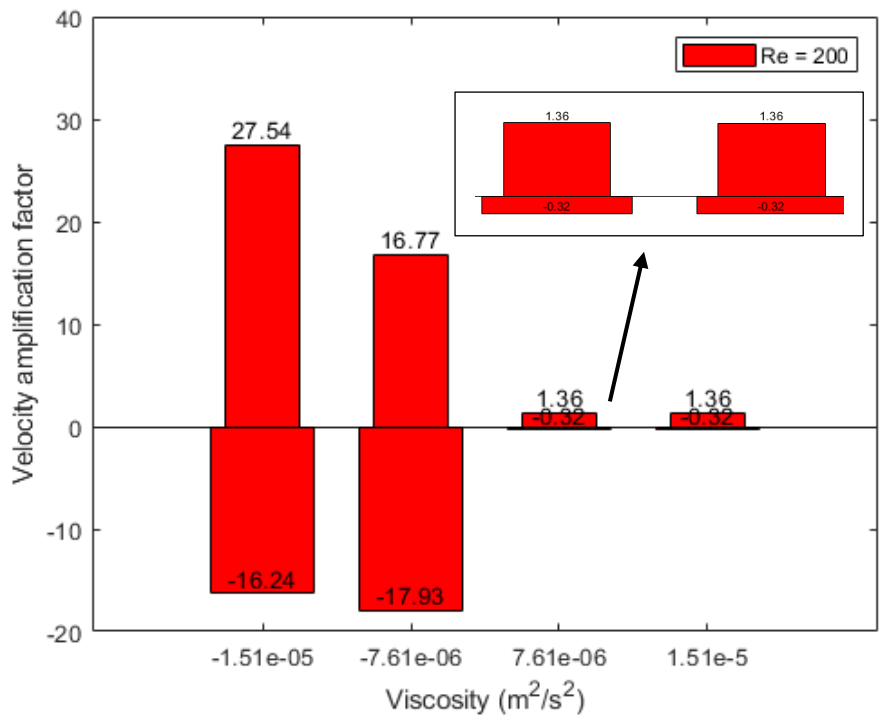

Email: sreetambhaduri@aol.com; ORCID: 0000-0002-5201-3976; Scopus ID: 57201682251

Fig.16. Comparison of VA factor across cross section at 9 mm distance from center of cylinder w.r.t. the inlet velocity for different viscosities at Re = 200

*Variation of Drag coefficient:*

Analyzing a flow past cylinder to study the effects of negative viscosity, the drag coefficient is a highly important parameter to study. The Fig.17 shows the variation and comparison of drag coefficient for different viscosities. Initially up to 3s of flow time the drag coefficient decreased due to flow separation never occurred for $\zeta = 1.51E - 05 \ m^2/s^2$ (Fig.17a). However, after 3s of flow time, the drag coefficient increased due to the occurrence of flow separation (Fig.17a). Flow separation never occurred at all for $\zeta = 7.61E - 06 \ m^2/s^2$ due to reduced viscosity (Fig.17b) than that of Fig.17(a). The negative viscosity shows an astonishing variation of drag coefficient. The drag coefficient highly fluctuates from both positive as well as in negative sides and the amplitude is very high as shown in Fig.17(c and d). This happened because of the viscous stimulation effect for the negative viscosities (Fig.17c and d). Due to increasing magnitude of negative viscosity from $\zeta = -7.61E - 06 \ m^2/s^2$ to $\zeta = -1.51E - 05 \ m^2/s^2$, the amplitude of variation of drag coefficient increased (Fig.17c and d). The Fig.17(e) shows the comparison of time averaged drag coefficient for different viscosities. Due to existence of flow separation, the time averaged drag coefficient is 66.67% higher for $\zeta = 1.51E - 05 \ m^2/s^2$ than that of for $\zeta = 7.61E - 06 \ m^2/s^2$. However, the time averaged drag coefficient turned out to be negative for the viscosity $\zeta = -7.61E - 06 \ m^2/s^2$ due to viscous stimulation effect. The time averaged drag coefficient is positive and 144 times higher than that of $\zeta = 1.51E - 05 \ m^2/s^2$ as a result of viscous stimulation effect instead of viscous dissipation effect.

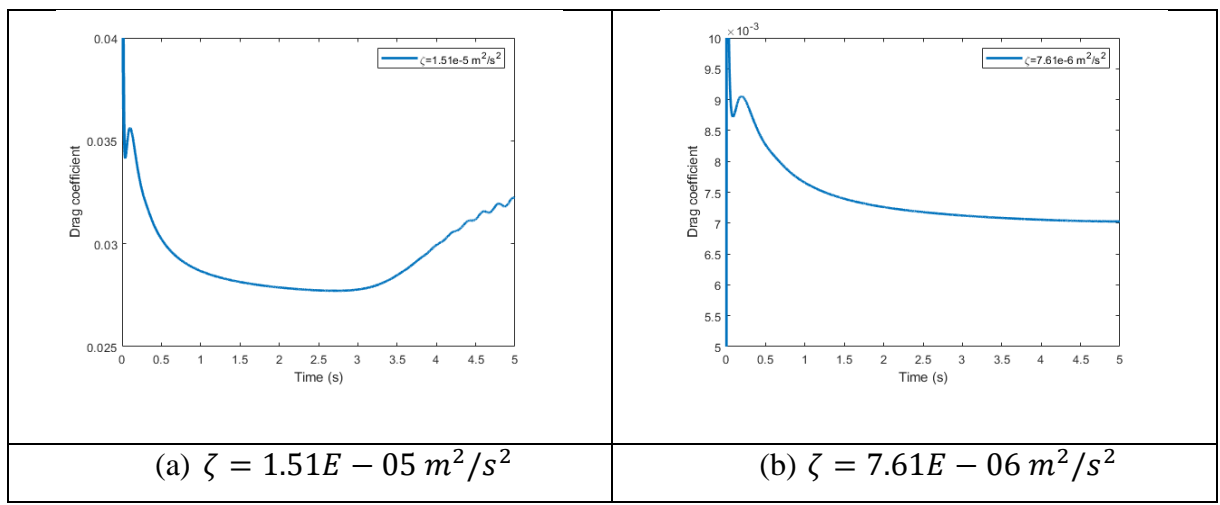

| (a) $\zeta = 1.51E - 05 \ m^2/s^2$ | (b) $\zeta = 7.61E - 06 \ m^2/s^2$ |

Email: sreetambhaduri@aol.com; ORCID: 0000-0002-5201-3976; Scopus ID: 57201682251

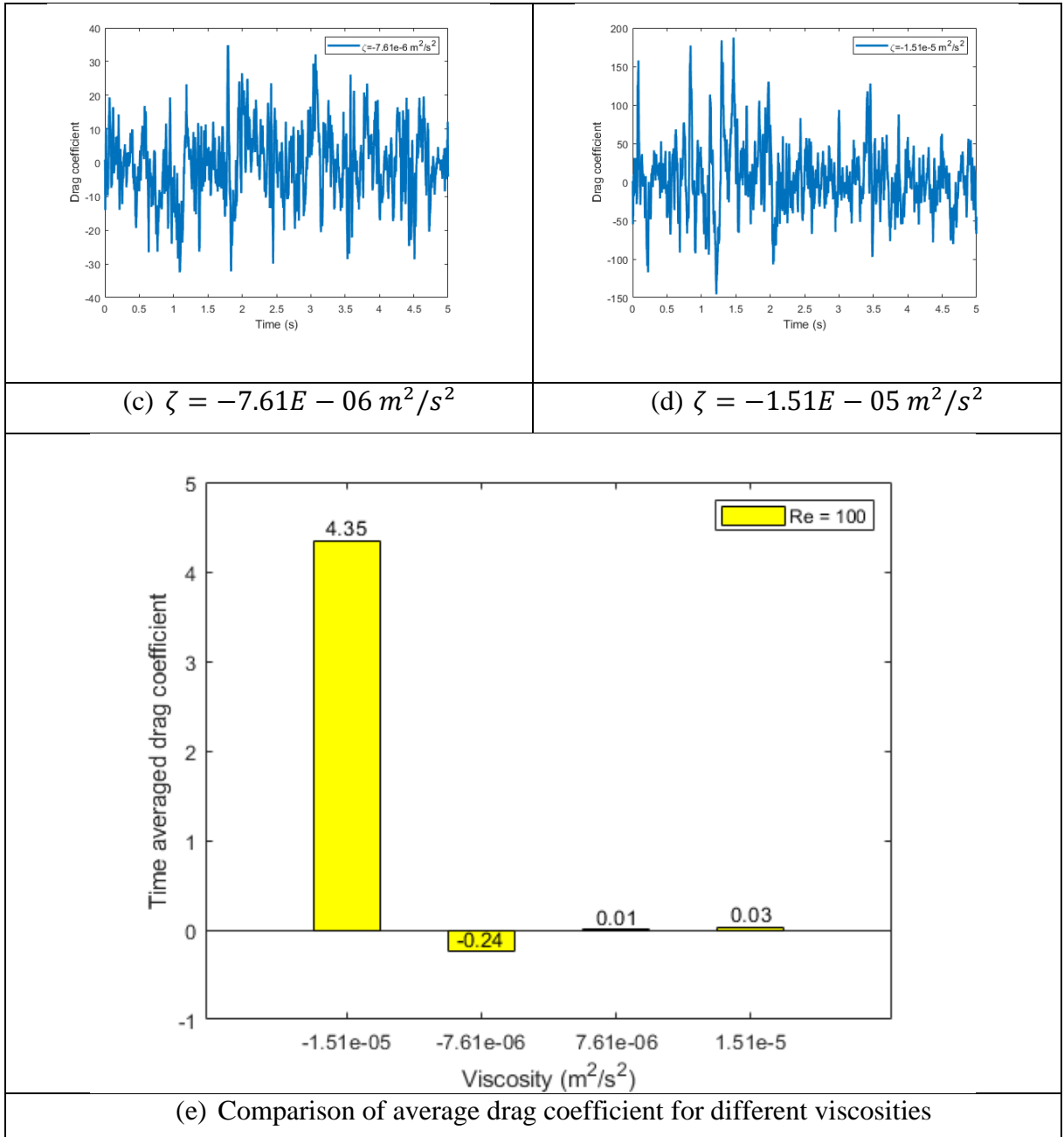

(c) $\zeta = -7.61E - 06 \ m^2/s^2$

(d) $\zeta = -1.51E - 05 \ m^2/s^2$

(e) Comparison of average drag coefficient for different viscosities

Fig.17. Variation and comparison of drag coefficients for different viscosities for inlet Re = 100

The Fig.18 shows the variation and comparison of drag coefficient for different viscosities. Initially up to 1.5s of flow time the drag coefficient decreased due to flow separation never occurred for $\zeta = 1.51E - 05 \ m^2/s^2$ (Fig.18a). However, after 1.5s of flow time, the drag coefficient increased due to the occurrence of flow separation (Fig.18a). Flow separation never occurred until 2.5s of flow time for $\zeta = 7.61E - 06 \ m^2/s^2$ due to reduced viscosity (Fig.18b) than that of Fig.18(a). The drag coefficient highly fluctuates from both positive as well as in negative sides and the amplitude is very high as shown in Fig.18(c and d). This happened

*Email: sreetambhaduri@aol.com; ORCID: 0000-0002-5201-3976; Scopus ID: 57201682251*

because of the viscous stimulation effect for the negative viscosities (Fig.18c and d). Due to increasing magnitude of negative viscosity from $\zeta = -7.61E - 06 \ m^2/s^2$ to $\zeta = -1.51E - 05 \ m^2/s^2$, the amplitude of variation of drag coefficient increased (Fig.18c and d). The Fig.18(e) shows the comparison of time averaged drag coefficient for different viscosities. Due to greater flow separation for $\zeta = 1.51E - 05 \ m^2/s^2$, the time averaged drag coefficient is 71.42% higher for $\zeta = 1.51E - 05 \ m^2/s^2$ than that of for $\zeta = 7.61E - 06 \ m^2/s^2$. However, the time averaged drag coefficient turned out to be negative for the viscosity $\zeta = -7.61E - 06 \ m^2/s^2$ due to viscous stimulation effect. The time averaged drag coefficient is positive and 132.4 times higher than that of $\zeta = 1.51E - 05 \ m^2/s^2$ due to viscous stimulation effect.

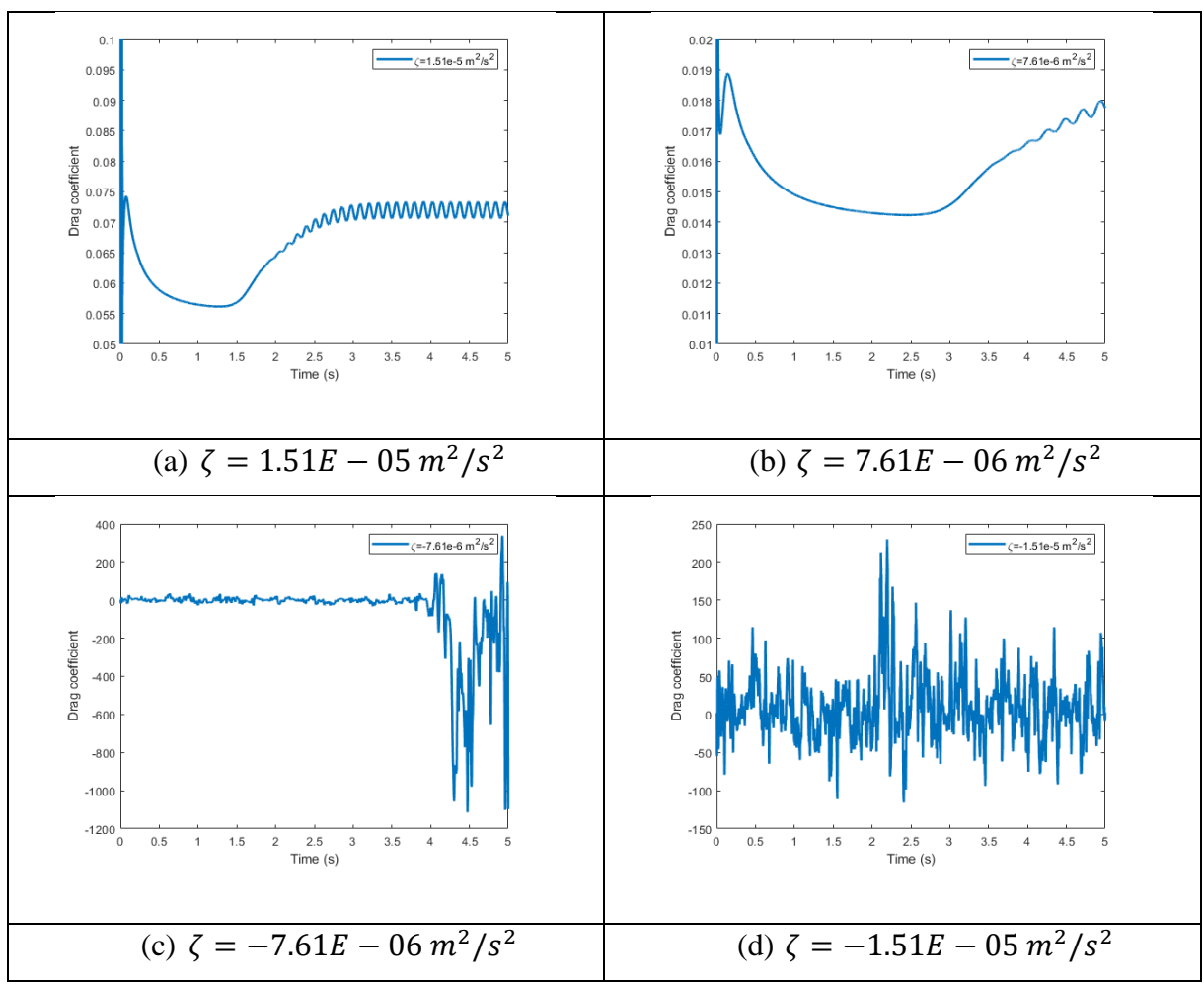

| (a) $\zeta = 1.51E - 05 \ m^2/s^2$ | (b) $\zeta = 7.61E - 06 \ m^2/s^2$ |
| (c) $\zeta = -7.61E - 06 \ m^2/s^2$ | (d) $\zeta = -1.51E - 05 \ m^2/s^2$ |

Email: sreetambhaduri@aol.com; ORCID: 0000-0002-5201-3976; Scopus ID: 57201682251

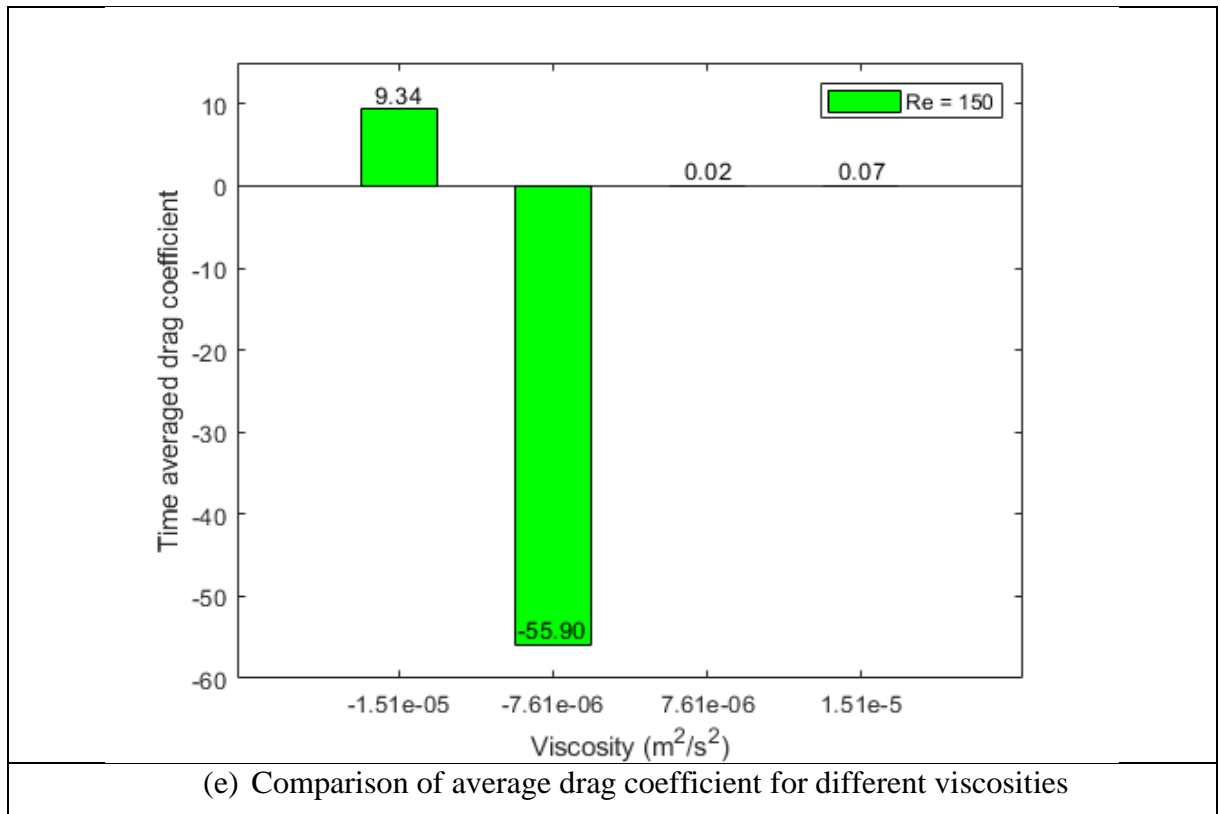

(e) Comparison of average drag coefficient for different viscosities

Fig.18. Variation and comparison of drag coefficients for different viscosities for inlet Re = 150

The Fig.19 shows the variation and comparison of drag coefficient for different viscosities. Initially up to 1s of flow time the drag coefficient decreased due to flow separation never occurred for $\zeta = 1.51E - 05 \; m^2/s^2$ (Fig.19a). However, after 1s of flow time, the drag coefficient increased due to the occurrence of flow separation (Fig.19a). Flow separation never occurred until 1.5s of flow time for $\zeta = 7.61E - 06 \; m^2/s^2$ due to reduced viscosity (Fig.19b) than that of Fig.19(a). The drag coefficient highly fluctuates from both positive as well as in negative sides and the amplitude is very high as shown in Fig.19(c and d). This happened because of the viscous stimulation effect for the negative viscosities (Fig.19c and d). Due to increasing magnitude of negative viscosity from $\zeta = -7.61E - 06 \; m^2/s^2$ to $\zeta = -1.51E - 05 \; m^2/s^2$, the amplitude of variation of drag coefficient increased (Fig.19c and d). The Fig.19(e) shows the comparison of time averaged drag coefficient for different viscosities. Due to greater flow separation for $\zeta = 1.51E - 05 \; m^2/s^2$, the time averaged drag coefficient is 75% higher for $\zeta = 1.51E - 05 \; m^2/s^2$ than that of for $\zeta = 7.61E - 06 \; m^2/s^2$. However, the time averaged drag coefficient turned out to be negative for the viscosity $\zeta = -7.61E - 06 \; m^2/s^2$ due to viscous stimulation effect. The time averaged drag coefficient is positive and 2589.5 times higher than that of $\zeta = 1.51E - 05 \; m^2/s^2$ due to viscous simulation effect.

Email: sreetambhaduri@aol.com; ORCID: 0000-0002-5201-3976; Scopus ID: 57201682251

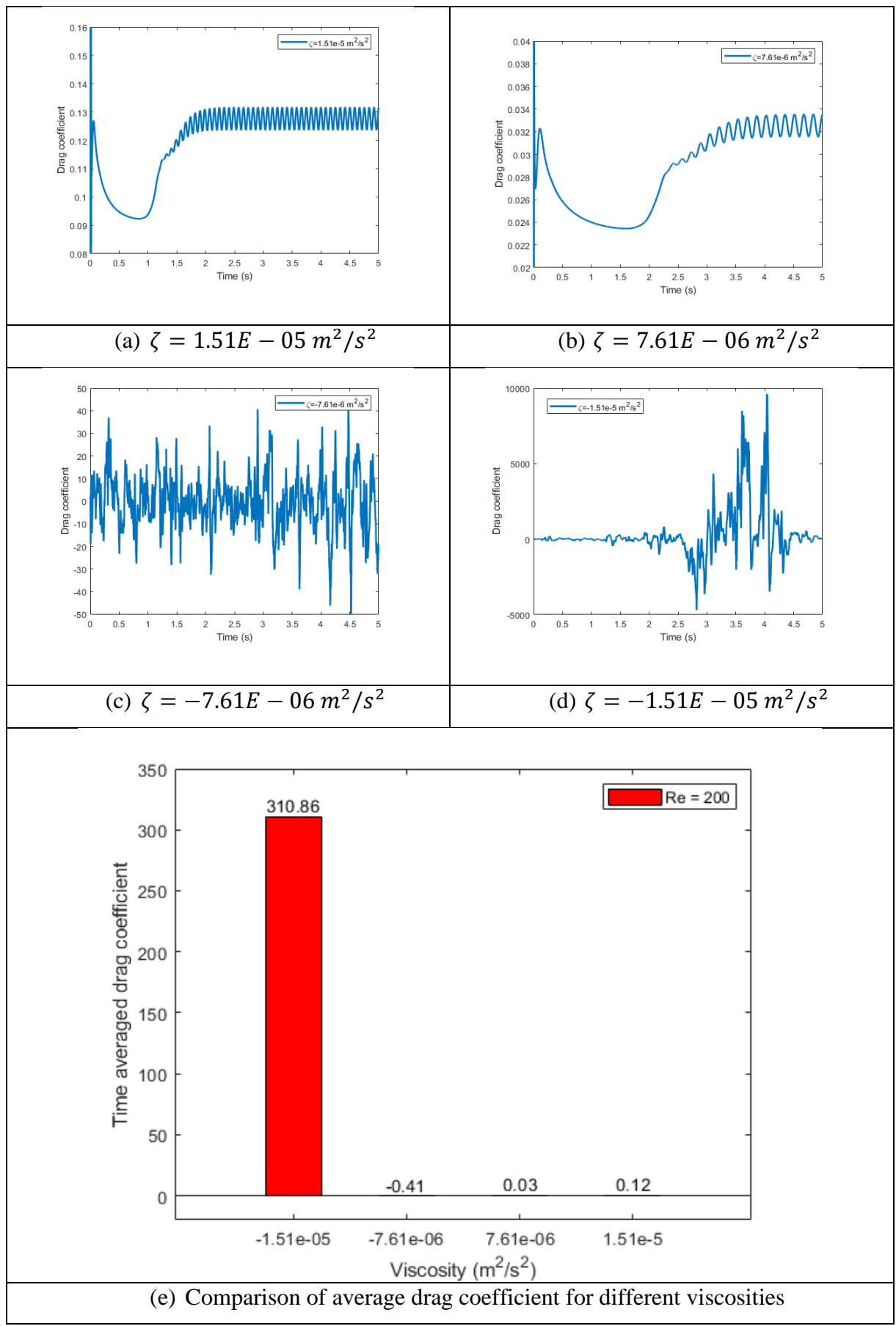

(a) $\zeta = 1.51E - 05\ m^2/s^2$

(b) $\zeta = 7.61E - 06\ m^2/s^2$

(c) $\zeta = -7.61E - 06\ m^2/s^2$

(d) $\zeta = -1.51E - 05\ m^2/s^2$

(e) Comparison of average drag coefficient for different viscosities

Email: sreetambhaduri@aol.com; ORCID: 0000-0002-5201-3976; Scopus ID: 57201682251

Fig.19. Variation and comparison of drag coefficients for different viscosities for inlet Re = 200

**Conclusion:**

The above discussion analyzed various aspects of the effect of negative viscosity compared to that of the positive viscosity in a fluid flow. The following points can be concluded from the discussion.

1. According to the second law of thermodynamics a fluid cannot have a negative viscosity unless and until the system develop a positive pressure difference between the inlet and outlet.
2. The positive viscosity induces a viscous dissipation effect which tends to slow down the motion of a fluid element. However, the negative viscosity develops a viscous stimulation effect which tends to accelerate the motion of a fluid element. But due to irregular motion of fluid elements, more disturbances get induced in the flow, which amplifies the disturbance formed by the flow past cylinder.
3. The inclusion of negative viscosity in general develops more disturbances inside the flow. The vorticity magnitude increased around 231 times and around 156 times for the inclusion of negative pairs of the positive viscosities w.r.t. that of $\zeta = 1.51E - 05 \ m^2/s^2$ and $\zeta = 7.61E - 06 \ m^2/s^2$ respectively when the inlet Re = 100. The vorticity magnitude increased around 130.8 times and around 1266.3 times for the inclusion of negative pairs of the positive viscosities w.r.t. that of $\zeta = 1.51E - 05 \ m^2/s^2$ and $\zeta = 7.61E - 06 \ m^2/s^2$ respectively when the inlet Re = 150. The vorticity magnitude increased around 204.2 times and around 231.5 times for the inclusion of negative pairs of the positive viscosities w.r.t. that of $\zeta = 1.51E - 05 \ m^2/s^2$ and $\zeta = 7.61E - 06 \ m^2/s^2$ respectively when the inlet Re = 200.
4. The number of CPs and IPs increases with increasing magnitude of negative viscosity due to increasing viscous stimulation effect. However, the number of CPs and IPs remains constant with increasing magnitude of positive viscosity.
5. The positive VA factor is around 28 times and 16 times higher for the negative pairs of positive viscosity w.r.t. that of $\zeta = 1.51E - 05 \ m^2/s^2$ and $\zeta = 7.61E - 06 \ m^2/s^2$ respectively for inlet Re = 100. However, the negative VA factor is around 37.2 times and 127.3 times higher for the negative pairs of positive viscosity w.r.t. that of $\zeta = 1.51E - 05 \ m^2/s^2$ and $\zeta = 7.61E - 06 \ m^2/s^2$ respectively for Re = 100. The

Email: sreetambhaduri@aol.com; ORCID: 0000-0002-5201-3976; Scopus ID: 57201682251

    positive VA factor is around 7.4 times and 202.6 times higher for the negative pairs of positive viscosity w.r.t. that of $\zeta = 1.51E - 05 \, m^2/s^2$ and $\zeta = 7.61E - 06 \, m^2/s^2$ respectively for inlet Re = 150. However, the negative VA factor is around 25.5 times and 2010.5 times higher for the negative pairs of positive viscosity w.r.t. that of $\zeta = 1.51E - 05 \, m^2/s^2$ and $\zeta = 7.61E - 06 \, m^2/s^2$ respectively for Re = 150. The positive VA factor is around 19.2 times and 11.3 times higher for the negative pairs of positive viscosity w.r.t. that of $\zeta = 1.51E - 05 \, m^2/s^2$ and $\zeta = 7.61E - 06 \, m^2/s^2$ respectively for inlet Re = 200. However, the negative VA factor is around 49.7 times and 55 times higher for the negative pairs of positive viscosity w.r.t. that of $\zeta = 1.51E - 05 \, m^2/s^2$ and $\zeta = 7.61E - 06 \, m^2/s^2$ respectively for Re = 200.

6. The inclusion of negative viscosity developed more large-scale disturbances, due to which the drag coefficient oscillated with a large amplitude. As a result, the drag coefficients for the positive viscosity are always positive, but the drag coefficients for the negative viscosity oscillates from both positive and negative magnitudes with a high amplitude.

7. The drag coefficient increased around 144 times and decreased around 23 times for the inclusion of negative pairs of the positive viscosities w.r.t. that of $\zeta = 1.51E - 05 \, m^2/s^2$ and $\zeta = 7.61E - 06 \, m^2/s^2$ respectively when the inlet Re = 100. The drag coefficient increased around 132.4 times and decreased around 2794 times for the inclusion of negative pairs of the positive viscosities w.r.t. that of $\zeta = 1.51E - 05 \, m^2/s^2$ and $\zeta = 7.61E - 06 \, m^2/s^2$ respectively when the inlet Re = 150. The drag coefficient increased around 2589.5 times and decreased around 12.6 times for the inclusion of negative pairs of the positive viscosities w.r.t. that of $\zeta = 1.51E - 05 \, m^2/s^2$ and $\zeta = 7.61E - 06 \, m^2/s^2$ respectively when the inlet Re = 200.

**Nomenclature:**

| Symbols | Names of parameter | Symbols | Names of parameter |
|---|---|---|---|
| $\dot{W}_{sh}$ | Rate of shaft work | $\rho$ | Density |
| $\vec{u}$ | Velocity of flow | $p$ | Pressure |
| $\mu$ | Actual viscosity of fluid | $\vec{j}$ | Current density |
| $\vec{B}$ | Magnetic field | $\theta$ | Angle |
| $\zeta$ | Effective viscosity | $\vec{\nabla}$ | $(\frac{\partial}{\partial x}\hat{\imath} + \frac{\partial}{\partial y}\hat{\jmath} + \frac{\partial}{\partial z}\hat{k})$ |

Email: sreetambhaduri@aol.com; ORCID: 0000-0002-5201-3976; Scopus ID: 57201682251

| $\vec{\omega}$ | Vorticity magnitude or Angular Velocity | $\dot{F}_l$ | Rate of flow work |
|---|---|---|---|
| $\Delta KE$ | Kinetic energy | $\dot{m}$ | Mass flow rate |
| $\Delta PE$ | Potential energy | $t$ | Time |
| $\dot{F}_{em}$ | Rate of Lorentz force | $\dot{S}_u$ | Rate of entropy generation for the universe |
| $\Delta s$ | Change of entropy of the system | $T$ | Temperature of the system in absolute scale |
| $h$ | Enthalpy of the system | $A_c$ | Surface area of the cylinder |
| $L$ | Thickness of the flow domain | Re | Reynold's number |
| VA | Velocity amplification | $V_x$ | X direction velocity |

**Acknowledgement:**

The author shows gratitude to SIMFLOW Technologies for providing free license to support this work.

**Disclaimer:**

This work has been performed from January, 2021 to March, 2021. All the simulation data has been acquired from the author's personal computing facility.

**References:**

1. Yunus, A. C. (2010). Fluid Mechanics: Fundamentals And Applications (Si Units). Tata McGraw Hill Education Private Limited.
2. Balescu, R., Prigogine, I., & Giovannini, A. (2000). Equilibrium statistical mechanics (pp. 119-122). World scientific.
3. Landau, L. D., & Lifshitz, E. M. (1987). Course of Theoretical Physics, Vol. 6: Fluid Mechanics 2, nd, Ed.
4. Sivashinsky, G., & Yakhot, V. (1985). Negative viscosity effect in large-scale flows. The Physics of fluids, 28(4), 1040-1042.
5. L. D. Meshalkin and Ya. G. Sinai, J. Appl. Math. Mech. (PMM) 25, 1700 (1961).
6. Gama, S., Vergassola, M., & Frisch, U. (1995). Negative Isotropic Eddy Viscosity: A Common Phenomenon in two Dimensions. In Navier—Stokes Equations and Related Nonlinear Problems (pp. 351-355). Springer, Boston, MA.

*Email:* sreetambhaduri@aol.com; ORCID: 0000-0002-5201-3976; Scopus ID: 57201682251

7. König, M., Noack, B. R., & Eckelmann, H. (1993). Discrete shedding modes in the von Karman vortex street. Physics of Fluids A: Fluid Dynamics, 5(7), 1846-1848.

8. Pal, R. (2019). Teach Second Law of Thermodynamics via Analysis of Flow through Packed Beds and Consolidated Porous Media. Fluids, 4(3), 116.

9. Cassel, K. W., & Conlisk, A. T. (2014). Unsteady separation in vortex-induced boundary layers. Philosophical Transactions of the Royal Society A: Mathematical, Physical and Engineering Sciences, 372(2020), 20130348.

10. Williams III, J. C. (1977). Incompressible boundary-layer separation. Annual Review of Fluid Mechanics, 9(1), 113-144.

11. Ma, H., & Duan, Z. (2020). Similarities of Flow and Heat Transfer around a Circular Cylinder. Symmetry, 12(4), 658.

12. Baratta, M., Chiriches, S., Goel, P., & Misul, D. (2020). CFD modelling of natural gas combustion in IC engines under different EGR dilution and H2-doping conditions. Transportation Engineering, 2, 100018.

13. Liu, Q., Gómez, F., Perez, J. M., & Theofilis, V. (2016). Instability and sensitivity analysis of flows using OpenFOAM®. Chinese Journal of Aeronautics, 29(2), 316-325.

14. Lee, S. B. (2017). A study on temporal accuracy of OpenFOAM. International Journal of Naval Architecture and Ocean Engineering, 9(4), 429-438.

15. Zhao, W., Zou, L., Wan, D., & Hu, Z. (2018). Numerical investigation of vortex-induced motions of a paired-column semi-submersible in currents. Ocean Engineering, 164, 272-283.

16. Ye, S., Lin, Y., Xu, L., & Wu, J. (2020). Improving Initial Guess for the Iterative Solution of Linear Equation Systems in Incompressible Flow. Mathematics, 8(1), 119.

17. Devals, C., Zhang, Y., Dompierre, J., Vu, T. C., Mangani, L., & Guibault, F. (2014, March). 3D casing-distributor analysis with a novel block coupled OpenFOAM solver for hydraulic design application. In IOP Conference Series: Earth and Environmental Science (Vol. 22, No. 2, p. 022005). IOP Publishing.

18. Dutykh, D. (2016). How to overcome the Courant-Friedrichs-Lewy condition of explicit discretizations?. arXiv preprint arXiv:1611.09646.